\documentstyle[11pt,twoside,pasp3D,epsf]{article}

\begin{document}

\title{High--Redshift Gravitationally Lensed Galaxies and Tunable Filter Imaging}

\author{Paul C. Hewett}
\affil{Institute of Astronomy, Madingley Road, Cambridge, CB3 0HA, UK}

\author{Stephen J. Warren}
\affil{Blackett Laboratory, Imperial College of Science Technology
and Medicine, Prince Consort Rd, London SW7 2BZ, UK} 

\author{Jon P. Willis}
\affil{Institute of Astronomy, Madingley Road, Cambridge, CB3 0HA, UK}

\author{Joss Bland--Hawthorn}
\affil{Anglo--Australian Observatory, PO Box 296, Epping, NSW 1710, Australia}

\author{Geraint F. Lewis}
\affil{Department of Physics and Astronomy, University of
Victoria, PO Box 3055, Victoria BC, V8W 3P7, Canada
and Astronomy Department, University of Washington, Box 351580,
Seattle, WA 98195, USA}

\begin{abstract}
Most surveys for multiply--imaged gravitational lenses, outside of rich
galaxy clusters, are based on sifting through large samples of
distant sources to identify the rare examples of lensing. An
alternative strategy, based on the selection of optimal
lines--of--sight, offers a number of significant advantages. Utilising
the multiplex capability of wide--area multifibre spectroscopy, together
with tunable Fabry--Perot imaging, it is now possible to undertake such
investigations. Progress in compiling a sample of high--redshift
star--forming galaxies, gravitationally lensed by massive early--type
galaxies at intermediate redshift, $z\sim 0.4$, is described.
\end{abstract}

\section{Introduction}

The potential of gravitational lenses to increase our knowledge
concerning the amount and nature of dark matter, constrain key
cosmological parameters and offer uniquely detailed views of faint
distant objects is well known (Blandford and Narayan 1992). The
development over the last decade of the field of ``weak lensing'' in
which statistical studies of many sources, only slightly perturbed by
the effects of lensing, are made has produced important new results
concerning the distribution of mass in galaxy clusters and very large
scale structures. However, the assembly of examples of systems
multiply imaged by individual galaxies has proved extremely hard.

Searches for individual examples of strong lensing have relied on the
examination of a sample of objects, such as quasars or flat--spectrum
radio--sources, where a large fraction of the sample lie at high
redshift. Thus, towards each object there is a significant
path--length over which an intervening deflector may interpose itself
close to the line--of--sight. The lens search proceeds through the
identification of sources whose morphology, multiple images or
extended arcs for example, is consistent with the effects of
lensing. Further imaging, at different wavelengths, and spectroscopy
is then necessary to establish the source as a bona--fide lens and to
obtain redshifts for the source and the deflecting galaxy. In practice,
obtaining the redshifts is very difficult, particularly for
radio--selected objects, and in the compilation of Kochanek et al
(1999: {\it http://cfa-www.harvard.edu/castles}) only 19 of the 45
lensed systems possess both deflector and source redshifts.

An alternative search strategy is to examine optimal lines--of--sight
by identifying a population of very effective deflectors, where it
is known that any source lying behind the deflector will be
significantly lensed, and then to examine the spectra of the deflectors
for evidence of lensed background sources. Miralda--Escud\'{e} and
Leh\'{a}r (1992) pointed out that provided the surface density of
faint, small, galaxies at high redshift is large, significant numbers
of galaxy--galaxy lenses should exist. Subsequent observational
developments have shown that the surface density of high--redshift,
star--forming objects is indeed large (e.g., Steidel et al. 1996, Hu,
Cowie and McMahon 1998).  Provided a suitable sample of deflectors can
be identified the optimal line--of--sight search strategy offers
significant advantages, including i) high efficiency, the probability a
lens will be seen along a line--of--sight is significant, ii) the
deflector and source redshifts may be readily acquired, allowing the
full lensing geometry to be defined, iii) the small, but extended,
star--forming objects lead to resolved gravitational lenses, not unlike
the radio--rings arising from morphologically extended radio emission,
which provide much greater constraints on the deflector masses than the
more familiar two-- or four--image lenses of unresolved quasars.

Using APM measures of United Kingdom Schmidt Telescope $B_JRI$ plates
it is possible to identify the ideal population of deflectors
\---\ massive, bulge--dominated, galaxies at redshift $z\sim 0.4$,
essentially half-way between ourselves and any high redshift source.
Specifically, locating the population of relatively bright, $m_R\le
20$, red, $B_J-R \ge 2.2$, galaxies with redshifts $0.25 \le z \le 0.6$
is straightforward (Warren et al.  1996).  The galaxy population has a
surface density of $\sim 50\,$deg$^{-2}$ and associated with each
galaxy there is an area of sky, $\sim 1\,$arcsec$^2$, in which any
distant source will be multiply imaged, with an associated increase in
brightness of a factor $\ga 10$. These early--type galaxies represent
essentially optimal lines--of--sight to search for examples of strong
lensing.

The presence of a lens is revealed by the detection of an anomalous
emission line in the spectrum of one of the target distant early-type
galaxies, so obtaining spectra of a large sample of the deflector
galaxies represents the first stage in the lens survey.  Examination
of intermediate--resolution optical spectra of an initial sample of
160 colour--selected early--type galaxies revealed the presence of an
emission line at $5589$\AA \ in a galaxy with redshift
$z=0.485$. Follow--up spectroscopy (Warren et al.  1998) and imaging
(Warren et al. 1999) have confirmed the B0047--2808 system as an
optical Einstein ring with the source, a star--forming galaxy at
$z=3.595$, the first confirmed example of a normal galaxy lensing
another normal galaxy and a demonstration of the viability of the
optimal line--of--sight survey strategy.

\section{The Spectroscopic Survey}

\begin{figure}
\plotone{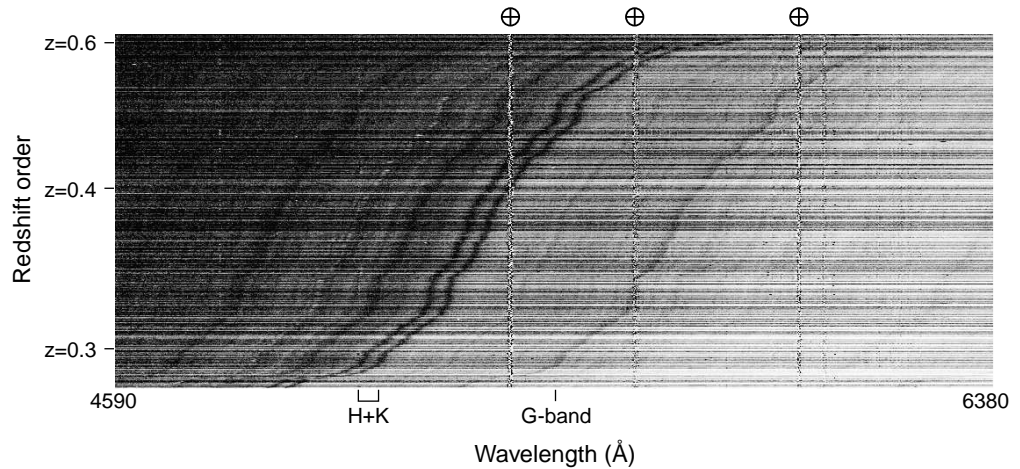}
\caption{Observed--frame spectra of 403 distant early-type galaxies,
redshifts $0.3<z<0.6$, arranged by increasing redshift. A number of
prominent night--sky features are visible as vertical lines while
features present in the galaxies, such as Calcium H+K and the G--band
move to longer wavelength with increasing redshift.} \label{fig-1}
\end{figure}

With an efficient method for acquiring spectra along many optimal
lines--of--sight there is the prospect of obtaining a large sample,
10--20 objects, of spatially resolved gravitationally lensed systems.
The low--surface density of the galaxies on the sky means the
Anglo--Australian Telescope's 2dF multifibre instrument, with a
$3\,{\rm deg}^2$ field, is ideally suited to the initial spectroscopy.
Total exposure times of $\sim 8000\,$s produce galaxy spectra for
which the completeness of redshift measurement is $95\%$ and in which
anomalous emission lines of fluxes $\sim 5\times 10^{-17}\,$erg
s$^{-1}$cm$^{-2}$ may be reliably detected i.e. fluxes comparable to
those seen in high--redshift galaxy samples (e.g., Hu et al. 1998) can
be reached. The unlensed fluxes are a factor $\sim 10$ fainter.
Example spectra of galaxies from 2dF are shown in
Figure~\ref{fig-1}. Two fields, producing $\sim 250$ galaxy spectra,
can be observed per night.

\begin{figure}
\plotfiddle{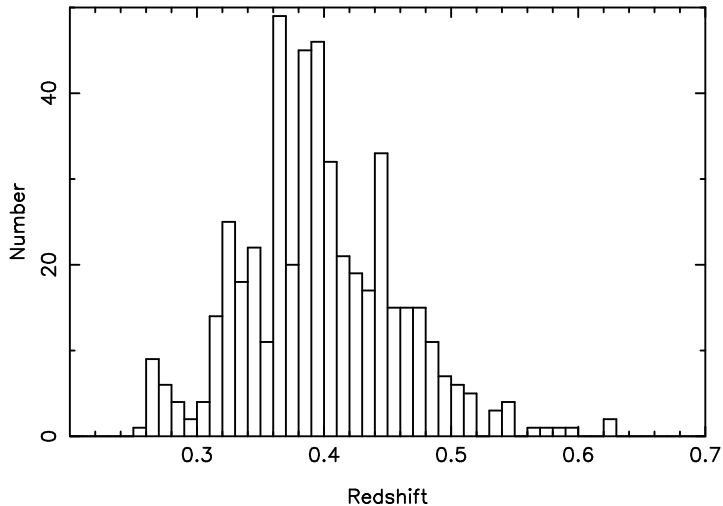}{6.0cm}{270}{40}{40}{-150}{+220}
\caption{Number--redshift histogram for the 485 galaxies observed over
two--nights at the AAT in 1998 September using the 2dF instrument.} \label{fig-2}
\end{figure}
Figure~\ref{fig-2}
shows the number--redshift distribution for all 485 galaxies observed
in a two--night 2dF run. The median redshift is $z_{med}=0.391$ and the
sample consists of luminous, relatively high--redshift objects that
barely feature in wide--angle, bright surveys (e.g. Colless 1998) or
narrow--angle, deep surveys (e.g. Cohen et al. 1999; Figure~1b).

\section{Tunable Filter Imaging}

Follow--up of an emission line detection in the discovery spectra
would, until recently, have involved the acquisition of a second
slit--spectrum to confirm the line. Subsequent narrow--band ($\sim
30$\AA) imaging, at an essentially random wavelength, to investigate
the morphology of the emission, necessitates the fabrication of a
custom filter, which is both time consuming and costly. However, the
availability of tunable Fabry--Perot imaging, specifically the Taurus
Tunable Filter (TTF) instruments at the AAT and William Herschel
Telescope, enable monochromatic images to be obtained for emission
lines at wavelengths covering virtually the whole optical spectrum.
A resolving power of up to $\sim 1000$ allows the signal--to--noise ratio to
be maximised by using narrow bandpasses, $\sim 10$\AA, well matched to
the narrow emission features. Observing efficiency has been further
improved with the recently commissioned broad--narrow shuffle mode,
that allows a comparison broadband image to be obtained during the
observation with only a small increase in the total exposure time.  The
sensitivity of the TTF system is such that confirmation of emission at
the target wavelength can be achieved as quickly as with a slit
spectrum. Indeed, the imaging  is superior in that even with a
relatively wide slit it is possible to exclude prominent emission
components, which may lie anywhere on a ring up to 3 arcsec in
diameter. Thus, spectroscopic follow--up may be dispensed with entirely
and confirmation and narrow--band imaging achieved using the TTF.

\begin{figure}
\plotfiddle{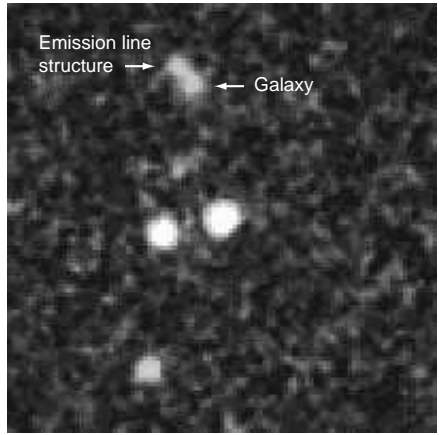}{7.0cm}{0}{80}{80}{-250}{-220}
\caption{Greyscale image of a $37 \times 37\,$arcsec region containing
the gravitational lens candidate B0042--2752 \---\ the double image visible towards the top. North is up and East to the left.}\label{fig-3}
\end{figure}

Figure~\ref{fig-3} 
illustrates the effectiveness of the TTF. The image is a composite of a
broad--band $J$ exposure, from UKIRT, and an $1800\,$s TTF exposure
from the AAT. The candidate lens system is the double image seen at
top--centre. The lower of the two components is the core of a galaxy,
redshift $z=0.519$, magnitude $m_R=18.8$, coordinates 00$^h$ 42$^m$
49.5$^s$ -27$^\circ$ 52$^\prime$ 17$^{\prime\prime}$ (Equinox B1950.0).
The spectrum is that of a normal early--type galaxy with the addition
of a weak emission line at $5800$\AA. The upper of the components
visible in Figure~\ref{fig-3}, separated by $\sim 2\,$arcsec from the
centre of the galaxy, is visible only in the TTF exposure, which
covered a $\sim 10$\AA \ bandpass centred at $5800$\AA. The detection
of the component in the narrow--band image is unambiguous. Combining
the image with two additional TTF exposures produces an image where
some evidence of extended emission may be present but much deeper TTF
observations are required to confirm any lower surface brightness
extended structure or the presence of fainter lensed images.

The most probable identification of the emission is with Ly$\alpha$
$1216$\AA \ from a star--forming galaxy at redshift $z=3.77$. With this
identification B0042--2752 is a very similar system to B0047--2808, the
first lens discovered in the survey. The deflector and source in
B0042--2752 are at slightly higher redshifts and the observed emission
line flux is a factor $\sim 5$ lower, $f_{\lambda} \sim 1 \times
10^{-16}\,{\rm erg}\,{\rm cm}^{-2}\,{\rm s}^{-1}$.  However, without
confirmation of the source redshift, via identification of a second
emission line in the infrared ([O {\footnotesize III}] 5007 would
appear at $2.388\,\mu$), or evidence for a morphology unambiguously
that of a lens, B0042-2752 remains only a lens candidate. Alternative
explanations are certainly not particularly attractive. If the emission
at $5800$\AA \ is [O {\footnotesize II}] 3727 the velocity of the
line--producing region relative to the galaxy is $\sim +11,000\,{\rm
km\,s}^{-1}$, ruling out any plausible physical association, even within
a rich cluster of galaxies. Chance projection of a faint
H\,{\footnotesize II} galaxy is possible, with $z_{em}=0.556$ if the
line is [O\,{\footnotesize II}] 3727, or $z_{em}=0.158$ if the line is
[O\,{\footnotesize III}] 5007.  There is no indication of any continuum
emission from broad--band imaging data, or the TTF off--band images,
and the putative H\,{\footnotesize II} galaxy must possess a large line
equivalent--width, with an intrinsically faint continuum. The object
would, for example, lie in the sparsely populated upper portion of
Figure~2 of Hogg et al. (1998).

The survey spectroscopic observations are sensitive to the presence of
an emission line object in an area of $\sim 5\,$arcsec$^2$ per galaxy,
giving a surveyed area of $\sim 800\,$arcsec$^2$ for the sample of 160
galaxies. Assuming, rather optimistically, that [O\,{\footnotesize II}]
or [O\,{\footnotesize III}] emission from a star--forming galaxy would
be detected in our spectra over the redshift range $0.1 \le z \le 0.8$
gives a total volume of $\sim 22\,h^{-3}\,{\rm Mpc}^3$
($H_0=100\,h\,{\rm km\,s}^{-1}$, $q_0=0.5$, $\Lambda=0$) in which an
H\,{\footnotesize II} galaxy may be found. Therefore, the space density
of H\,{\footnotesize II} galaxies, with properties consistent with the
constraints outlined above, must be $\ga 0.01 h^3\,{\rm Mpc}^{-3}$ for
the {\it a priori} probability of finding such an object to be
significant. Such a space density is substantially above that found in
emission--line surveys to comparable line fluxes (e.g., Thompson,
Djorgovski and Trauger 1995).

Further observations will determine whether B0042-2752 is the second
lensed high--redshift galaxy to be identified in the survey.
Irrespective of the outcome the combination of the AAT 2dF instrument
and TTF imaging offers the prospect of compiling a well--defined
sample of resolved gravitational lenses with a homogeneous deflector
population. Investigation of the sample will provide unique
information on the mass--to--light ratio in field elliptical galaxies
and the morphology, emission line properties and masses of
high--redshift star--forming objects currently too faint to be studied
by any other means.

\acknowledgments

Bahram Mobasher kindly obtained the $J$--band image used to generate
Figure~\ref{fig-3}. GFL acknowledges support from the Pacific Institute
for Mathematical Sciences (1998-1999). The authors acknowledge the data
and analysis facilities provided by the Starlink Project which is run
by CCLRC on behalf of PPARC.

\end{document}